\documentclass[pre,amsmath,amssymb,superscriptaddress,showpacs]{revtex4}
\usepackage{graphicx,bm}
\begin{document}

\title{Coherent structures in an electron beam}
\author{A. Ifliberi} 
\affiliation{INFM Milano Universit\`a, INFN Sezione di Milano,
Dipartimento di Fisica, Universit\`a degli Studi di Milano, Milano, Italy}
\author{G. Bettega}  
\affiliation{INFM Milano Universit\`a, INFN Sezione di Milano,
Dipartimento di Fisica, Universit\`a degli Studi di Milano, Milano, Italy}
\author{M. Cavenago}
\affiliation{INFN Laboratori Nazionali di Legnaro, Legnaro, Italy}
\author{F. De Luca}
\affiliation{INFM Milano Universit\`a, INFN Sezione di Milano,
Dipartimento di Fisica, Universit\`a degli Studi di Milano, Milano, Italy}
\author{R. Pozzoli} 
\affiliation{INFM Milano Universit\`a, INFN Sezione di Milano,
Dipartimento di Fisica, Universit\`a degli Studi di Milano, Milano, Italy}
\author{M. Rom\'e} 
\affiliation{INFM Milano Universit\`a, INFN Sezione di Milano,
Dipartimento di Fisica, Universit\`a degli Studi di Milano, Milano, Italy}
\author{Yu.~Tsidulko}
\affiliation{Budker Institute of Nuclear Physics, Novosibirsk, 
Russian Federation}

\begin{abstract}

The formation and evolution of coherent structures in a low-energy electron beam produced in a Malmberg-Penning trap is investigated by means of  CCD diagnostics. The electrons are emitted from a thermionic cathode and their energy is controlled by an acceleration grid. By varying the spatial distribution of the energy of emitted electrons, different space charge effects are observed, as, e. g., a sharp or a gradual transition to a space charge dominated regime. The variation of the coherent structures along the beam is studied by varying the electron density or/and the value of the confined magnetic field. The observed processes are interpreted using a tridimensional particle-in-cell code which solves the Vlasov-Poisson system in zeroth order drift approximation.

\end{abstract}

\maketitle

\section{Introduction}

Recently the investigation of single species plasmas has 
received considerable attention owing to its wide range of applications 
and its connection with basic problems of fluid dynamics and statistical 
physics \cite{davidson,rmp99}. 
To study the evolution of such systems cylindrical (Malmberg-Penning) traps are commonly used: the transverse confinement is provided by a uniform magnetic field, and the axial confinement by an electrostatic potential well. In the case of electron plasmas the cyclotron period is usually the shortest time scale in the system and the drift approximation of the electron motion is valid. It
has been found that when the bounce period of the axial motion 
is much shorter than the period of azimuthal drift, 
the axially averaged dynamics is properly described by the two-dimensional 
(2D) Euler (or drift-Poisson) system \cite{degrassie77}

\begin{equation}
\frac{\partial{n}}{\partial{t}}+{\bf v_{\perp}}  \cdot \nabla_{\perp}n =0, 
 \label{continuity0}
 \end{equation}
 \begin{equation}
 {\bf v_{\perp}}=-(c/B) \nabla_{\perp}  \phi \times  {\bf e}_z, 
  \label{drift0}
 \end{equation}
 \begin{equation}
\nabla_{\perp}^2\phi= 4 \pi e n.
       \label{potential0}
 \end{equation}
 
 In that limit, the electric potential $\phi$ is proportional to the stream function, and the plasma vorticity  $\zeta = {\bf e}_z \cdot \nabla \times {\bf v_{\perp}}$ is proportional to $n/B$, being $n$ the (bounce averaged) plasma density and ${\bf e}_z$ the unit vector parallel to ${\bf B}$ (and to the axis of the device).
The formation and evolution of coherent structures in such conditions, including filamentation, merger, relaxation of turbulence, has been extensively studied using the inject-hold-dump technique, and good agreement with 2D numerical investigations has been found (see e.g. \cite{{crystal},{schecter99},{rome},{amoretti},{bettega04a}}).  
 On the long time scale (hundreds of the rotation period) the electron plasma can approach the equilibrium with a stable density profile with cylindrical symmetry, passing through
a series of long lasting states characterized by regular arrays of vortices (vortex crystals) \cite{crystal}.

\section{Structures in a low-energy electron beam}
The above limit cannot be applied to the case of a beam in the trap, where the electrons continuously flow from the emitting source to the charge collector (phosphor screen), held at a fixed potential (few kV). 
In this case, the formation of structures and relevant phenomena can develop on the time scale of electron transit through the beam. 
The nonlinear dynamics of the space-charge-limited flow \cite{luginsland02} 
is strongly affected by the presence of the axial magnetic field: 
when reflection occurs in the central part of the beam a hollow 
electron column forms and fast coherent structures arise 
\cite{apl1}, possibly due to the development of diocotron 
instability.

The mentioned complex behaviour both of trapped 
plasmas and beams can be described in the framework of the zeroth order drift 
approximation. 
Assuming the guiding centers
distribution $f({\bf{x}}, {\bf{v}},t)=
F(v_{\|},{\bf{x}},t)\delta({\bf{v}}_{\bot} - {\bf{v}}_{E})$,  where $v_\|$ denotes the velocity component parallel
to the magnetic field and ${\bf{v}}_{\bot}$ the perpendicular velocity, the relevant Vlasov-Poisson system reads:
\begin{equation}
\begin{split}
& \frac{\partial F}{\partial t}+ ({\bf{v}}_{E} + v_\|{\bf e}_z ) \cdot \nabla F + \frac{e}{m} \ {\bf e}_z 
\cdot \nabla \phi  \frac{\partial F}{\partial v_{\parallel}}  = 0, \\
& \nabla^2\phi = 4 \pi e n \,, 
\end{split}
\label{mep3e1}
\end{equation}
where $n({\bf x} , t) = \int F \, d v_{\parallel} $ is the electron density and the electric drift ${\bf{v}}_{E}$ is given by Eq. (\ref{drift0}).

\begin{figure}[h]
\centering
\includegraphics[scale=0.11]{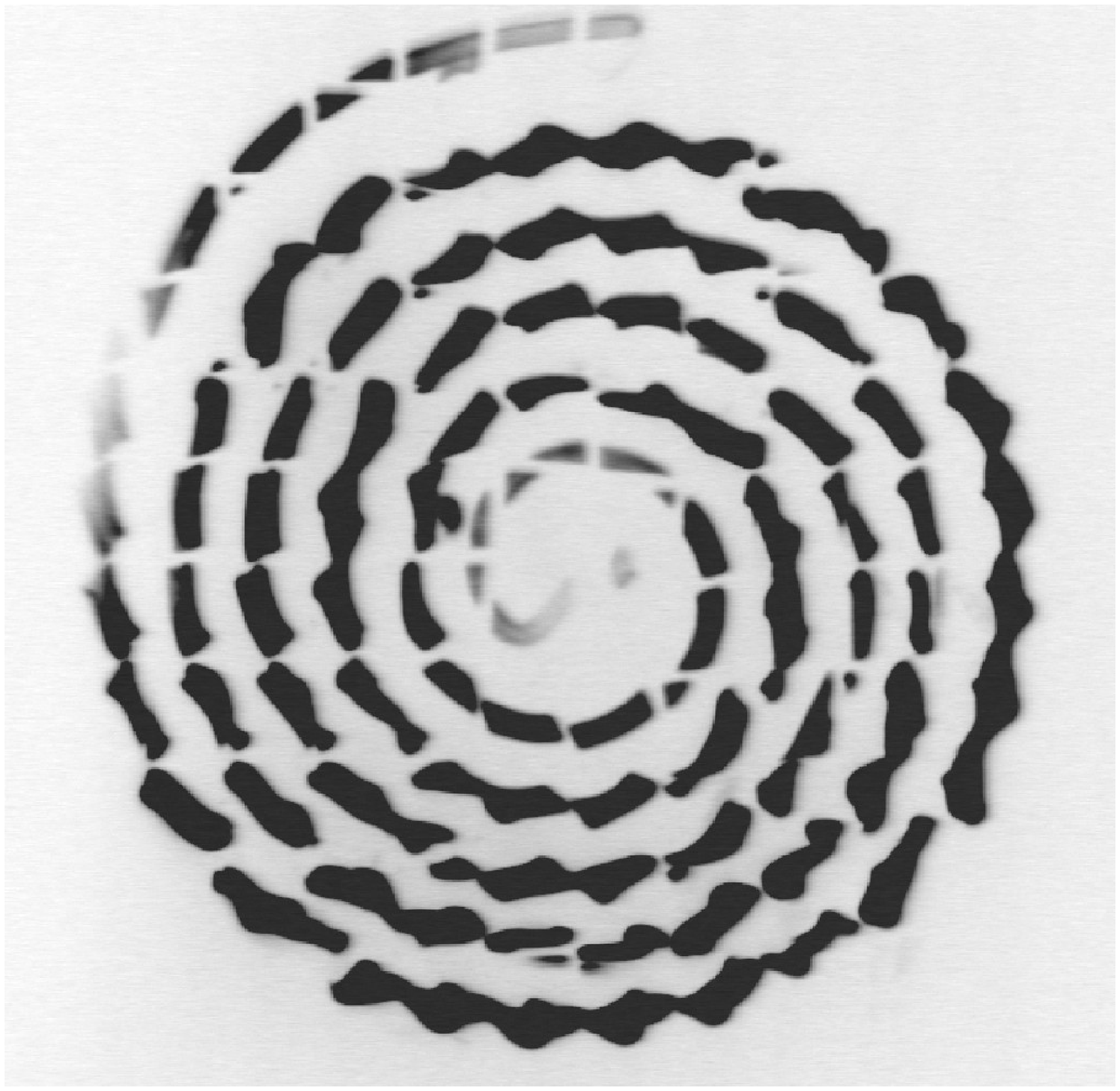}
\includegraphics[scale=0.11]{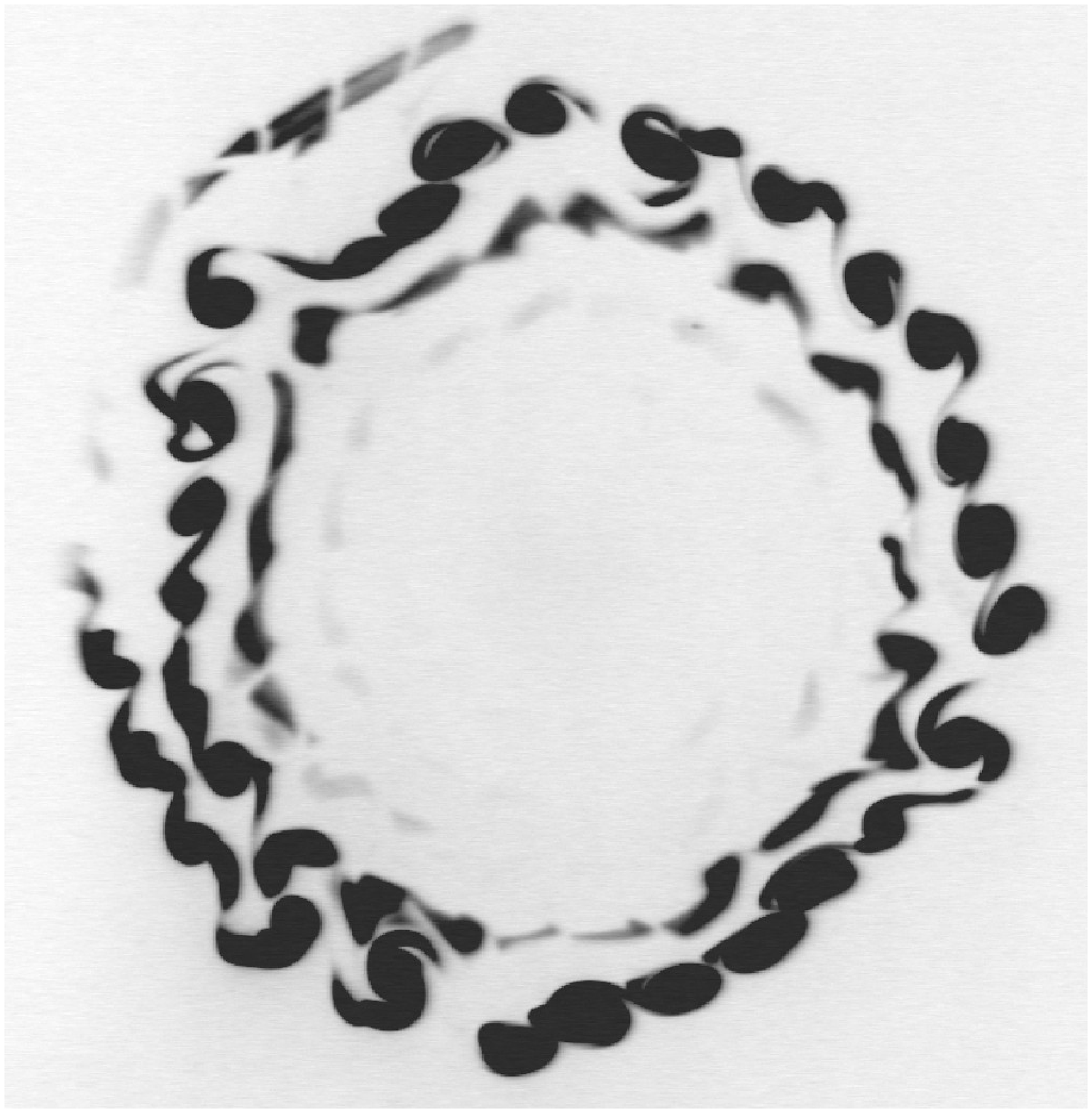}
\caption{\label{sharp}{Sharp transition to a space charge dominated regime, induced by increasing the emission current of the cathode.}}
\end{figure}

A rapid development of the structures is observed when, by increasing the emission current of the cathode, a sharp transition to a space charge dominated regime occurs \cite{apl1}. The pictures in Fig.\ref{sharp} show the variation of the beam density distribution at the phosphor screen, observed in two conditions close to the transition point. The difference between the emission currents in the two pictures is less than one percent: passing from the left picture to the right, an annular beam is produced, thus indicating the appearance around the axis of the electron column of a region unaccessible to beam electrons. This sharp modification occurs when the energy of the electrons is enough high (about 20 eV).  If the energy of emitted electrons is about half of that value, or lower, a gradual transition to a space charge dominated regime occurs: as the emission current is increased, one by one the central rings of the source become less bright, until they disappear, while coherent structures form in the remaining beam.   

To account for the mechanism of such transitions 
it is to note that
stationary solutions with cylindrical symmetry, i.e. with $\phi, n, w$ independent of the polar angle $\theta$, and a purely azimuthal ${\bf v_{\perp}}$, may exist and obey the following equation:
\begin{equation}
   \nabla ^{2} \phi(r,z)= 4 \pi e   \frac {n_{0}(r)w_{0}(r)}{\sqrt {(2/m)(e\psi_{0}(r)+e \phi(r,z))}},
    \label{eqpoiss}
    \end{equation}
where  $e \psi_0= m {w_0}^2/2- e \phi_0$, and the quantities with index $0$ represent the values at $z=0$ (at the source grid, with $\phi_0=0$).
A typical behaviour of the solution of Eq. (\ref{eqpoiss}) in a space charge dominated case is characterized by the presence of the reflection surface $\psi_{0}(r)+\phi(r,z)=0$ which encloses a spatial region not accessible by the electrons emitted from the source. Close to this surface  the plasma density exhibits large values. The behavior of the solution is related to the problem of space-charge-limited emission, which has been recently revisited in the literature \cite{{lau01}, {akimov01}, {luginsland02}}.  

To account for the fast development of the coherent structures observed on the phosphor screen in the space charge regime we note that, along most of the beam length $L$, close to the boundary of the hollow region a sharp radial density variation is present and the radial part of the Laplacian becomes dominant with respect to longitudinal: then, the plasma vorticity  becomes  proportional to plasma density, and the dynamics becomes 2D Eulerian. Thus, following the mechanism of diocotron instability coherent structures can easily develop. 
In the mentioned region the stationary continuity equation approximately reads: 
\begin{equation}
v_\| \frac {\partial}{\partial{z}}n+ {\bf v_{\perp}}  \cdot \nabla_{\perp}n =0,
    \label{continuity2}
    \end{equation}
where  $v_\|$ is almost constant, and $\phi$ obeys Eq. (\ref{potential0}).

Since the observed structures are tridimensional, to match the almost 2D (r,z) electron reflection due to space charge and the 2D (r, $\theta$) development of vortices we have to take into account the time evolution of the system, in which the boundary conditions at the source are dependent on $\theta$. To interpret the experimental results a 3D time dependent problem has to be solved, and can be approached numerically.
To this aim, a 3D PIC code has been developed \cite{tsidulko03} which solves the Vlasov-Poisson system (\ref{mep3e1}), both for trapped plasma and beam configurations. Examples of the solutions, computed in a space charge dominated regime are reported in Ref.\cite{rome04}.  The obtained results confirm the overall picture given above. In that regime a part of the plasma near the axis is reflected back to the grid while a annulus of high density is formed, and gives rise to the development of vortices. 

\begin{figure}[h]
\centering
\includegraphics[scale=0.22]{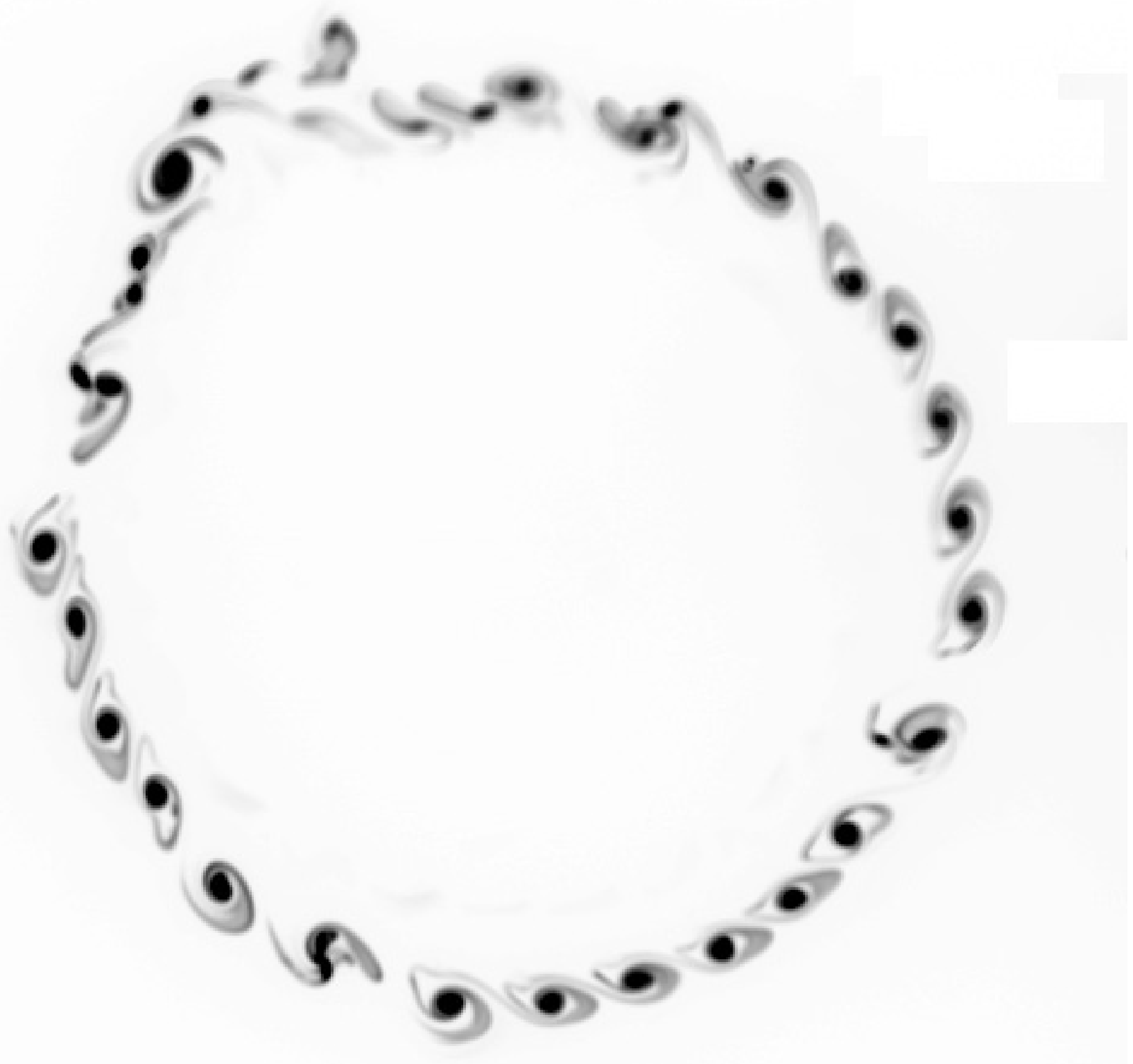}
\includegraphics[scale=0.22]{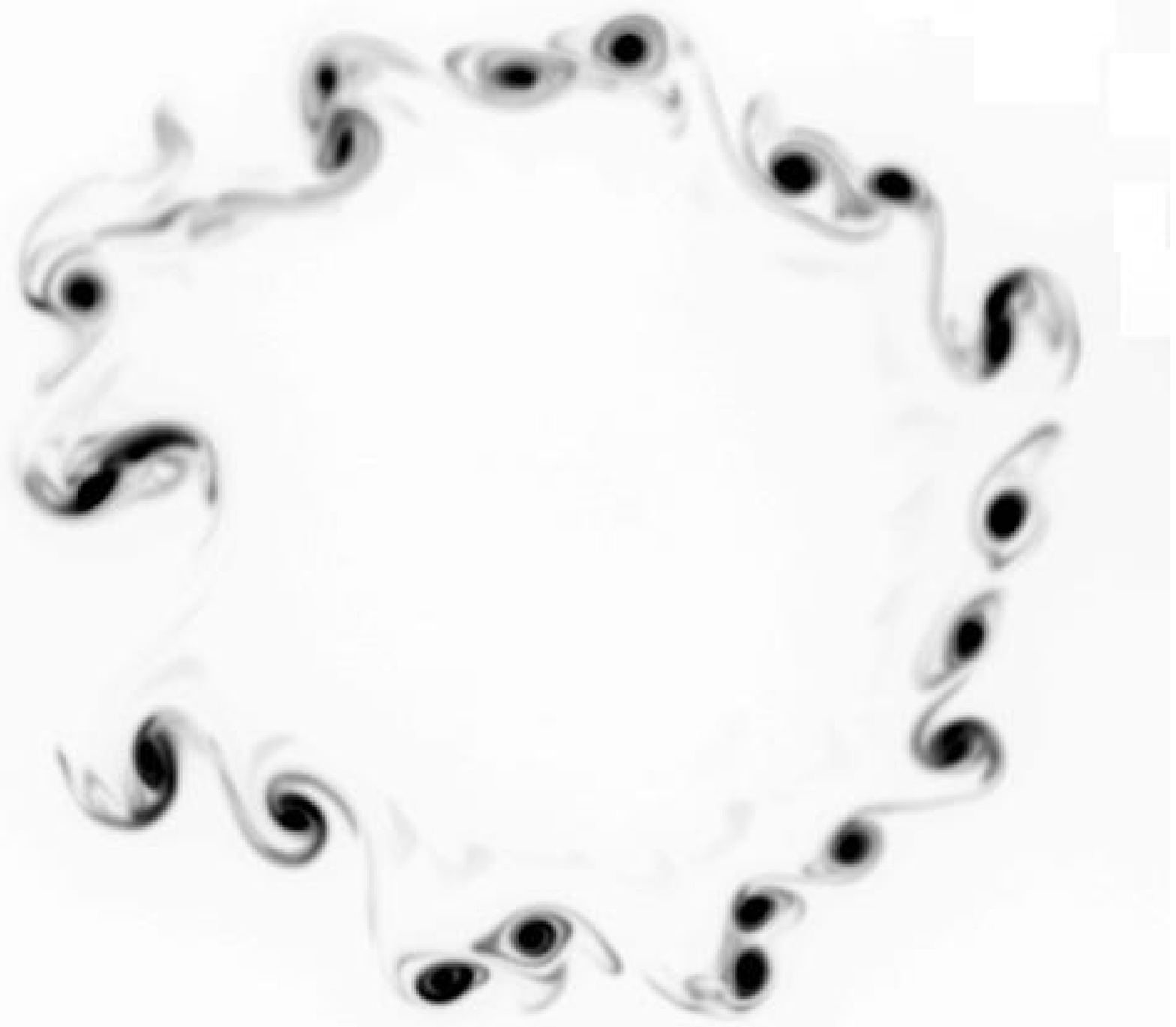}
\includegraphics[scale=0.22]{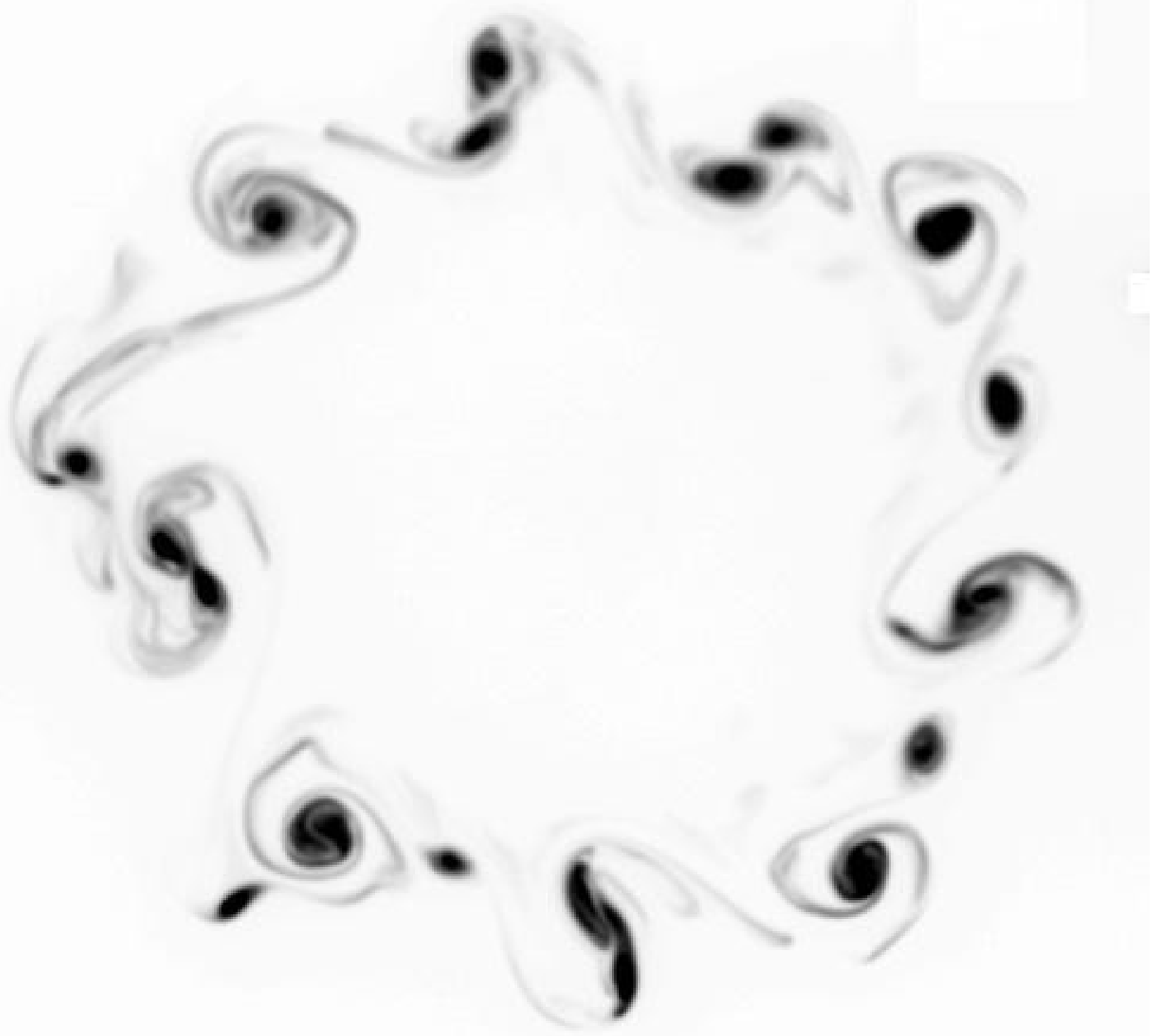}
\caption{\label{beam}Evolution of an electron beam in ELTRAP in the space charge regime. The images have been taken using different values of the confining magnetic field (decreasing from left to the right picture)}
\end{figure}

To study experimentally the transport of the structures along the beam, we can observe from Eq. (\ref{continuity2}) that the effective evolution time of the system is proportional to $L/B$. The evolution along the beam can be investigated by measuring the electron density on the phosphor screen at different $B$ values. A "time evolution" can be then obtained by slowly decreasing the magnetic field.
The observed density evolution on the phosphor screen is described by a purely 2D dynamics as far as the mentioned approximations are valid. Three images representing the density distribution of a beam on the phosphor screen at different times, obtained by decreasing $B$, are shown in Fig. \ref{beam}. 

\begin{figure}[h]
\centering
\includegraphics[scale=0.70]{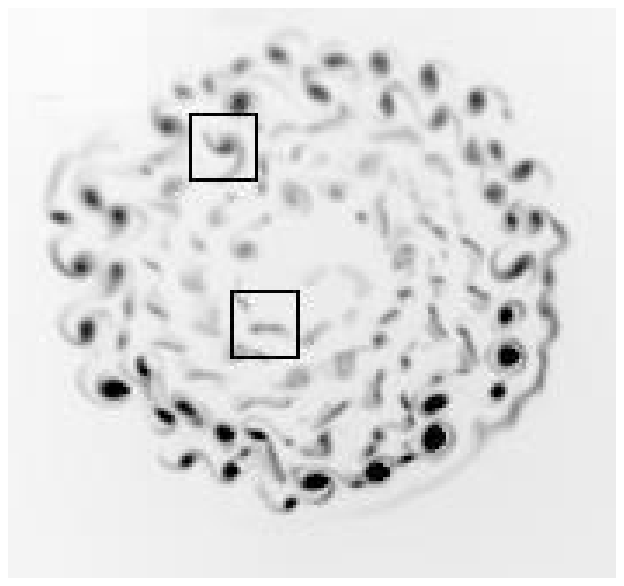}
\includegraphics[scale=0.70]{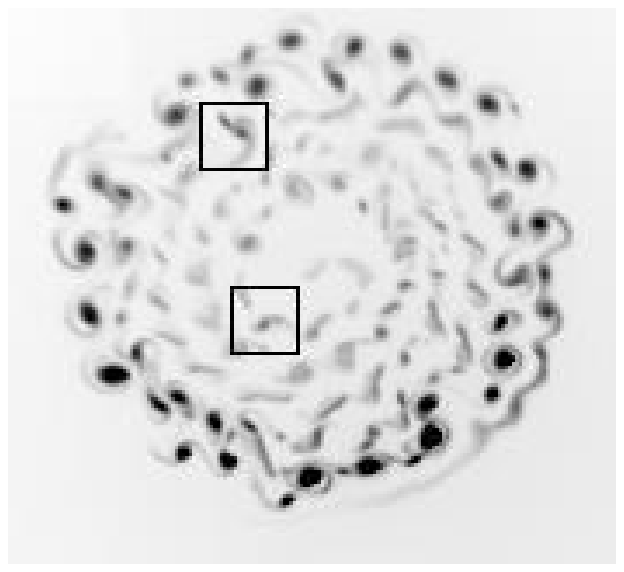}
\includegraphics[scale=0.70]{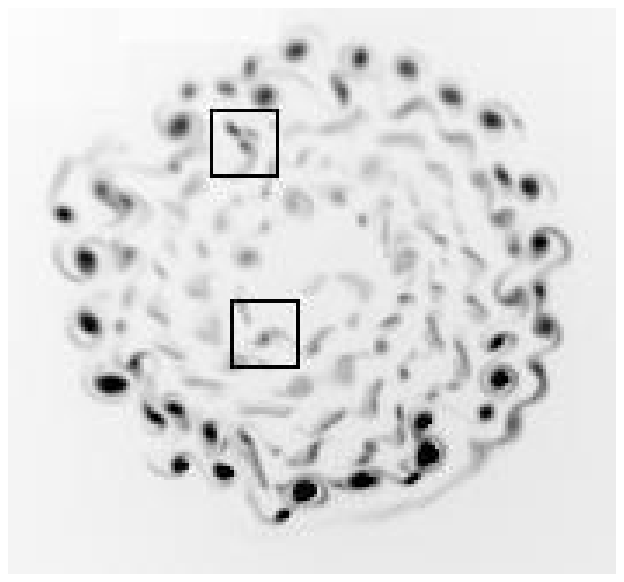}
\caption{\label{beam2} The squares point out two different evolutions of the coherent structures, due to a local vorticity variation.}
\end{figure}

Since in most of a hollow beam the plasma vorticity is proportional to the plasma density, different evolutions of the coherent structures can be observed in different beam regions at the same time, depending on the local variation of the density. The three images in Fig \ref{beam2} are obtained by increasing the emission current: a typical merger process occurs close to the center of the beam, where the local density is increasing, while a  "inverse merger" process can be observed in the peripheral region, where the density is decreasing.

\section{Conclusions}
The formation and evolution of tridimensional coherent structures in a low-energy electron beam, where the space charge effects are dominant, have been investigated experimentally in a Malmberg-Penning trap, using CCD diagnostics. The main control parameters are the spatial distributions, at the source, of the electron density, energy and current, and the magnetic field. The reflection process has been investigated  by varying the electron energy and emission current, and sharp or gradual transition to the space charge dominated regime have been found. The longitudinal structure of the beam has been studied by varying the magnetic field or/and the emission current.  Some peculiar behaviors of vortex merger have been observed. It has been found that the observed behavior of the system, in the considered range of parameters, is in good agreement with the results obtained with a PIC code which solves the Vlasov-Poisson system in the zeroth order drift approximation.  

\bibliography{ILLIBERI}

\begin{thebibliography}{14}
\expandafter\ifx\csname natexlab\endcsname\relax\def\natexlab#1{#1}\fi
\expandafter\ifx\csname bibnamefont\endcsname\relax
  \def\bibnamefont#1{#1}\fi
\expandafter\ifx\csname bibfnamefont\endcsname\relax
  \def\bibfnamefont#1{#1}\fi
\expandafter\ifx\csname citenamefont\endcsname\relax
  \def\citenamefont#1{#1}\fi
\expandafter\ifx\csname url\endcsname\relax
  \def\url#1{\texttt{#1}}\fi
\expandafter\ifx\csname urlprefix\endcsname\relax\def\urlprefix{URL }\fi
\providecommand{\bibinfo}[2]{#2}
\providecommand{\eprint}[2][]{\url{#2}}

\bibitem[{\citenamefont{Davidson}(1990)}]{davidson}
\bibinfo{author}{\bibfnamefont{R.~C.} \bibnamefont{Davidson}},
  \emph{\bibinfo{title}{An Introduction to the Physics of Nonneutral Plasmas}}
  (\bibinfo{publisher}{Addison-Wesley}, \bibinfo{address}{Redwood City, USA},
  \bibinfo{year}{1990}).

\bibitem[{\citenamefont{Dubin and O'Neil}(1999)}]{rmp99}
\bibinfo{author}{\bibfnamefont{D.}~\bibnamefont{Dubin}} \bibnamefont{and}
  \bibinfo{author}{\bibfnamefont{T.}~\bibnamefont{O'Neil}},
  \bibinfo{journal}{Rev.\ Mod.\ Phys.} \textbf{\bibinfo{volume}{71}},
  \bibinfo{pages}{87} (\bibinfo{year}{1999}).

\bibitem[{\citenamefont{de~Grassie and Malmberg}(1977)}]{degrassie77}
\bibinfo{author}{\bibfnamefont{J.~S.} \bibnamefont{de~Grassie}}
  \bibnamefont{and} \bibinfo{author}{\bibfnamefont{J.~H.}
  \bibnamefont{Malmberg}}, \bibinfo{journal}{Phys.\ Rev.\ Lett.}
  \textbf{\bibinfo{volume}{39}}, \bibinfo{pages}{1077} (\bibinfo{year}{1977}).

\bibitem[{\citenamefont{Fine et~al.}(1995)\citenamefont{Fine, Cass, Flynn, and
  Driscoll}}]{crystal}
\bibinfo{author}{\bibfnamefont{K.~S.} \bibnamefont{Fine}},
  \bibinfo{author}{\bibfnamefont{A.~C.} \bibnamefont{Cass}},
  \bibinfo{author}{\bibfnamefont{W.~G.} \bibnamefont{Flynn}}, \bibnamefont{and}
  \bibinfo{author}{\bibfnamefont{C.~F.~.} \bibnamefont{Driscoll}},
  \bibinfo{journal}{Phys.\ Rev.\ Lett.} \textbf{\bibinfo{volume}{75}},
  \bibinfo{pages}{3277} (\bibinfo{year}{1995}).

\bibitem[{\citenamefont{Schecter et~al.}(1999)\citenamefont{Schecter, Dubin,
  Fine, and Driscoll}}]{schecter99}
\bibinfo{author}{\bibfnamefont{D.~A.} \bibnamefont{Schecter}},
  \bibinfo{author}{\bibfnamefont{D.~H.~E.} \bibnamefont{Dubin}},
  \bibinfo{author}{\bibfnamefont{K.~S.} \bibnamefont{Fine}}, \bibnamefont{and}
  \bibinfo{author}{\bibfnamefont{C.~F.} \bibnamefont{Driscoll}},
  \bibinfo{journal}{Phys.\ Fluids} \textbf{\bibinfo{volume}{11}},
  \bibinfo{pages}{905} (\bibinfo{year}{1999}).

\bibitem[{\citenamefont{Rom\'e et~al.}(2000)\citenamefont{Rom\'e, Brunetti,
  Califano, Pegoraro, and Pozzoli}}]{rome}
\bibinfo{author}{\bibfnamefont{M.}~\bibnamefont{Rom\'e}},
  \bibinfo{author}{\bibfnamefont{M.}~\bibnamefont{Brunetti}},
  \bibinfo{author}{\bibfnamefont{F.}~\bibnamefont{Califano}},
  \bibinfo{author}{\bibfnamefont{F.}~\bibnamefont{Pegoraro}}, \bibnamefont{and}
  \bibinfo{author}{\bibfnamefont{R.}~\bibnamefont{Pozzoli}},
  \bibinfo{journal}{Phys.\ Plasmas} \textbf{\bibinfo{volume}{7}},
  \bibinfo{pages}{2856} (\bibinfo{year}{2000}).

\bibitem[{\citenamefont{Amoretti et~al.}(2001)\citenamefont{Amoretti, Durkin,
  Fajans, Pozzoli, and Rom\'e}}]{amoretti}
\bibinfo{author}{\bibfnamefont{M.}~\bibnamefont{Amoretti}},
  \bibinfo{author}{\bibfnamefont{D.}~\bibnamefont{Durkin}},
  \bibinfo{author}{\bibfnamefont{J.}~\bibnamefont{Fajans}},
  \bibinfo{author}{\bibfnamefont{R.}~\bibnamefont{Pozzoli}}, \bibnamefont{and}
  \bibinfo{author}{\bibfnamefont{M.}~\bibnamefont{Rom\'e}},
  \bibinfo{journal}{Phys.\ Plasmas} \textbf{\bibinfo{volume}{8}},
  \bibinfo{pages}{3865} (\bibinfo{year}{2001}).

\bibitem[{\citenamefont{Bettega
  et~al.}(2004{\natexlab{a}})\citenamefont{Bettega, Cavaliere, Cavenago, {De
  Luca}, Illiberi, Kotelnikov, Pozzoli, {Rom\'e}, and Tsidulko}}]{bettega04a}
\bibinfo{author}{\bibfnamefont{G.}~\bibnamefont{Bettega}},
  \bibinfo{author}{\bibfnamefont{F.}~\bibnamefont{Cavaliere}},
  \bibinfo{author}{\bibfnamefont{M.}~\bibnamefont{Cavenago}},
  \bibinfo{author}{\bibfnamefont{F.}~\bibnamefont{{De Luca}}},
  \bibinfo{author}{\bibfnamefont{A.}~\bibnamefont{Illiberi}},
  \bibinfo{author}{\bibfnamefont{I.}~\bibnamefont{Kotelnikov}},
  \bibinfo{author}{\bibfnamefont{R.}~\bibnamefont{Pozzoli}},
  \bibinfo{author}{\bibfnamefont{M.}~\bibnamefont{{Rom\'e}}}, \bibnamefont{and}
  \bibinfo{author}{\bibfnamefont{Y.}~\bibnamefont{Tsidulko}}
  (\bibinfo{year}{2004}{\natexlab{a}}), \bibinfo{note}{{G. Bertin}, D. Farina
  and R. Pozzoli, editors, {\it {Plasmas in the Laboratory and in the
  Universe}}, AIP Conference Proceedings 703, page 48, American Institute of
  Physics}.

\bibitem[{\citenamefont{Luginsland et~al.}(2002)\citenamefont{Luginsland, Lau,
  Umstattd, and Watrous}}]{luginsland02}
\bibinfo{author}{\bibfnamefont{J.}~\bibnamefont{Luginsland}},
  \bibinfo{author}{\bibfnamefont{Y.}~\bibnamefont{Lau}},
  \bibinfo{author}{\bibfnamefont{R.}~\bibnamefont{Umstattd}}, \bibnamefont{and}
  \bibinfo{author}{\bibfnamefont{J.}~\bibnamefont{Watrous}},
  \bibinfo{journal}{Phys.\ Plasmas} \textbf{\bibinfo{volume}{9}},
  \bibinfo{pages}{2371} (\bibinfo{year}{2002}).

\bibitem[{\citenamefont{Bettega
  et~al.}(2004{\natexlab{b}})\citenamefont{Bettega, Cavaliere, Illiberi,
  Pozzoli, {Rom\'e}, Cavenago, and Tsidulko}}]{apl1}
\bibinfo{author}{\bibfnamefont{G.}~\bibnamefont{Bettega}},
  \bibinfo{author}{\bibfnamefont{F.}~\bibnamefont{Cavaliere}},
  \bibinfo{author}{\bibfnamefont{A.}~\bibnamefont{Illiberi}},
  \bibinfo{author}{\bibfnamefont{R.}~\bibnamefont{Pozzoli}},
  \bibinfo{author}{\bibfnamefont{M.}~\bibnamefont{{Rom\'e}}},
  \bibinfo{author}{\bibfnamefont{M.}~\bibnamefont{Cavenago}}, \bibnamefont{and}
  \bibinfo{author}{\bibfnamefont{Y.}~\bibnamefont{Tsidulko}},
  \bibinfo{journal}{Appl.\ Phys.\ Lett.} \textbf{\bibinfo{volume}{84}},
  \bibinfo{pages}{3807} (\bibinfo{year}{2004}{\natexlab{b}}).

\bibitem[{\citenamefont{Lau}(2001)}]{lau01}
\bibinfo{author}{\bibfnamefont{Y.}~\bibnamefont{Lau}}, \bibinfo{journal}{Phys.\
  Rev.\ Lett.} \textbf{\bibinfo{volume}{87}}, \bibinfo{pages}{278301}
  (\bibinfo{year}{2001}).

\bibitem[{\citenamefont{Akimov et~al.}(2001)\citenamefont{Akimov, Shamel,
  Kolinsky, Ender, and Kuznetsov}}]{akimov01}
\bibinfo{author}{\bibfnamefont{P.}~\bibnamefont{Akimov}},
  \bibinfo{author}{\bibfnamefont{H.}~\bibnamefont{Shamel}},
  \bibinfo{author}{\bibfnamefont{H.}~\bibnamefont{Kolinsky}},
  \bibinfo{author}{\bibfnamefont{A.}~\bibnamefont{Ender}}, \bibnamefont{and}
  \bibinfo{author}{\bibfnamefont{V.}~\bibnamefont{Kuznetsov}},
  \bibinfo{journal}{Phys.\ Plasmas} \textbf{\bibinfo{volume}{8}},
  \bibinfo{pages}{3788} (\bibinfo{year}{2001}).

\bibitem[{\citenamefont{Tsidulko et~al.}(2003)\citenamefont{Tsidulko, Pozzoli,
  and {Rom\'e}}}]{tsidulko03}
\bibinfo{author}{\bibfnamefont{Y.}~\bibnamefont{Tsidulko}},
  \bibinfo{author}{\bibfnamefont{R.}~\bibnamefont{Pozzoli}}, \bibnamefont{and}
  \bibinfo{author}{\bibfnamefont{M.}~\bibnamefont{{Rom\'e}}}
  (\bibinfo{year}{2003}), \bibinfo{note}{{M. Schauer}, T. Mitchell, R. Nebel,
  editors, {\it Non-Neutral Plasma Physics V}, AIP 692, page 279, Melville, New
  York, 2003, American Institute of Physics}.

\bibitem[{\citenamefont{{Rom\'e} et~al.}(2004)\citenamefont{{Rom\'e}, Pozzoli,
  Pravettoni, and Tsidulko}}]{rome04}
\bibinfo{author}{\bibfnamefont{M.}~\bibnamefont{{Rom\'e}}},
  \bibinfo{author}{\bibfnamefont{R.}~\bibnamefont{Pozzoli}},
  \bibinfo{author}{\bibfnamefont{M.}~\bibnamefont{Pravettoni}},
  \bibnamefont{and} \bibinfo{author}{\bibfnamefont{Y.}~\bibnamefont{Tsidulko}}
  (\bibinfo{year}{2004}), \bibinfo{note}{12th International Congress on Plasma
  Physics, 25-29 October 2004, Nice (France),
  http://hal.ccsd.cnrs.fr/ccsd-00001859}.

\end{thebibliography}

\end{document}